\documentclass [reprint,amsmath,amssymb,aps,superscriptaddress]{revtex4-2}
\usepackage{graphicx}
\usepackage{ulem}
\usepackage{subfigure}
\usepackage{verbatim}
\usepackage{dcolumn}
\usepackage{bm}
\usepackage{epsf}
\usepackage{xcolor}
\usepackage{hhline}
\usepackage{float}
\usepackage{enumerate}
\usepackage{bbm}
\usepackage{lipsum}

\usepackage{CJKutf8}
\usepackage{hyperref}
\hypersetup{
	colorlinks=true,   
	hypertexnames=false,
	linkcolor=blue,  
	citecolor=blue, 
	urlcolor =blue    
}

\begin{document}
	
	\preprint{APS/123-QED}
	
	\title{Microwave-to-optics conversion using magnetostatic modes and a tunable optical cavity}

	\author{Wei-Jiang Wu}
	\affiliation{Interdisciplinary Center of Quantum Information, State Key Laboratory of Extreme Photonics and Instrumentation, and Zhejiang Key Laboratory of Micro-nano Quantum Chips and Quantum Control, School of Physics, Zhejiang University, Hangzhou 310027, China}
	\author{Yi-Pu Wang}\email{yipuwang@zju.edu.cn}
	\affiliation{Interdisciplinary Center of Quantum Information, State Key Laboratory of Extreme Photonics and Instrumentation, and Zhejiang Key Laboratory of Micro-nano Quantum Chips and Quantum Control, School of Physics, Zhejiang University, Hangzhou 310027, China}
	\author{Jie Li}\email{jieli007@zju.edu.cn}
	\affiliation{Interdisciplinary Center of Quantum Information, State Key Laboratory of Extreme Photonics and Instrumentation, and Zhejiang Key Laboratory of Micro-nano Quantum Chips and Quantum Control, School of Physics, Zhejiang University, Hangzhou 310027, China}
	\author{Gang Li}
	\affiliation{State Key Laboratory of Quantum Optics and Quantum Optics Devices, and Institute of Opto-Electronics, Shanxi University, Taiyuan 030006, China}
	\author{J. Q. You}\email{jqyou@zju.edu.cn}
	\affiliation{Interdisciplinary Center of Quantum Information, State Key Laboratory of Extreme Photonics and Instrumentation, and Zhejiang Key Laboratory of Micro-nano Quantum Chips and Quantum Control, School of Physics, Zhejiang University, Hangzhou 310027, China}
	\affiliation{College of Optical Science and Engineering, Zhejiang University, Hangzhou 310027, China}

	\date{\today}

\begin{abstract}
		Quantum computing, quantum communication and quantum networks rely on hybrid quantum systems operating in different frequency ranges. For instance, the superconducting qubits work in the gigahertz range, while the optical photons used in communication are in the range of hundreds of terahertz. Due to the large frequency mismatch, achieving the direct coupling and information exchange between different information carriers is generally difficult. Accordingly, a quantum interface is demanded, which serves as a bridge to establish information linkage between different quantum systems operating at distinct frequencies. Recently, the magnon mode in ferromagnetic spin systems has received significant attention. While the inherent weak optomagnonic coupling strength restricts the microwave-to-optical photon conversion efficiency using magnons, the versatility of the magnon modes, together with their readily achievable strong coupling with other quantum systems, endow them with many distinct advantages. Here, we realize the magnon-based microwave-light interface by adopting an optical cavity with adjustable free spectrum range and different kinds of magnetostatic modes in two microwave cavity configurations. By optimizing the parameters, an internal conversion efficiency of $1.28\times10^{-7}$ is achieved. We analyze the impact of various parameters on the microwave-to-optics conversion. The study provides useful guidance and insights to further enhancing the microwave-to-optics conversion efficiency using magnons.
\end{abstract}
	\maketitle

\section{Introduction}
Quantum interfaces and transducers~\cite{RevModPhys.82.1041,PhysRevLett.109.130503,Zeuthen_2020}, which realize the coherent transformation of quantum information between different quantum systems, are necessary for building quantum networks~\cite{Kimble2008,RevModPhys.87.1379}. Different quantum systems function within specific frequency ranges. Superconducting qubit devices operating in the microwave frequency range are a promising platform for quantum information processing~\cite{doi:10.1146/annurev-conmatphys-031119-050605,Tsai}. Optical photons are more suited for long-distance quantum communication compared to microwave photons since they experience minimal loss and advanced technologies have been developed for detecting single optical photons~\cite{Gisin2007,https://doi.org/10.1002/qute.201900038,YUAN20101}. As a result, quantum interface and transducer schemes that facilitate conversion between microwave and optical photons by the aid of other systems are gaining increasing attention~\cite{lambertCoherentConversionMicrowave2020,hanMicrowaveopticalQuantumFrequency2021,fanSuperconductingCavityElectrooptics2018a,heaseBidirectionalElectroOpticWavelength2020a,higginbothamHarnessingElectroopticCorrelations2018,tuHighefficiencyCoherentMicrowavetooptics2022,welinskiElectronSpinCoherence2019,fernandez-gonzalvoCavityEnhancedRaman2019,evertsMicrowaveOpticalPhoton2019,chaiSinglesidebandMicrowavetoopticalConversion2022,zhuWaveguideCavityOptomagnonics2020,jiangEfficientBidirectionalPiezooptomechanical2020,ruedaElectroopticEntanglementSource2019,williamsonMagnetoOpticModulatorUnit2014a,jiangOpticallyHeraldedMicrowave2023,xuBidirectionalElectroopticConversion2021,sahuQuantumenabledOperationMicrowaveoptical2022}.

 In recent years, the optomagnonic microwave-to-optics transducer manifested as the magnon-induced light scattering has attracted much attention~\cite{Osada-16,zhangOptomagnonicWhisperingGallery2016,Haigh-16} owing to its unique advantages. For instance, it allows the magnon mode frequency to be tuned in a wide range and exhibits abundant nonlinear effects in ferrites~\cite{wangBistabilityCavityMagnon2018,bozhkoSupercurrentRoomtemperatureBose2016,shenLongTimeMemoryTernary2021,pirroAdvancesCoherentMagnonics2021,wuObservationNonlinearityHeatinginduced2022,shenMechanicalBistabilityKerrmodified2022a}, which  enrich the studies of the field. The ideal host for the magnon modes is the yttrium iron garnet (YIG) crystal, which is a ferrimagnetic material with a high density of spins~\cite{SpinWavesTheory2009a}. The high spin density makes the magnon mode easily achieve strong coupling with other physical degrees of freedom. The investigation of the magnon modes has predominantly resided within the classical realm for a lengthy time~\cite{pirroAdvancesCoherentMagnonics2021,Barman_2021}, mainly focusing on the understanding and manipulation of the dynamics of spin waves with a large number of magnons excited. However, the utilization of the macroscopic spin ensemble as a quantum interface is predicated upon its ability to exhibit quantum states~\cite{YUAN20221,Lachance-Quirion_2019}. Fortunately, this has now been actualized. The demonstrations of strong coupling between the superconducting qubit and YIG sphere~\cite{doi:10.1126/science.aaa3693,doi:10.1126/science.aaz9236}, magnon number states~\cite{doi:10.1126/sciadv.1603150}, quantum superposition states of a single magnon and the vacuum~\cite{PhysRevLett.130.193603}, and the Bell entangled state~\cite{xu2023deterministic} in the YIG-superconducting qubit hybrid system have served as a catalyst for the advancement of quantum magnonics.

\begin{figure*}[!t]
	\includegraphics[width=0.65\textwidth]{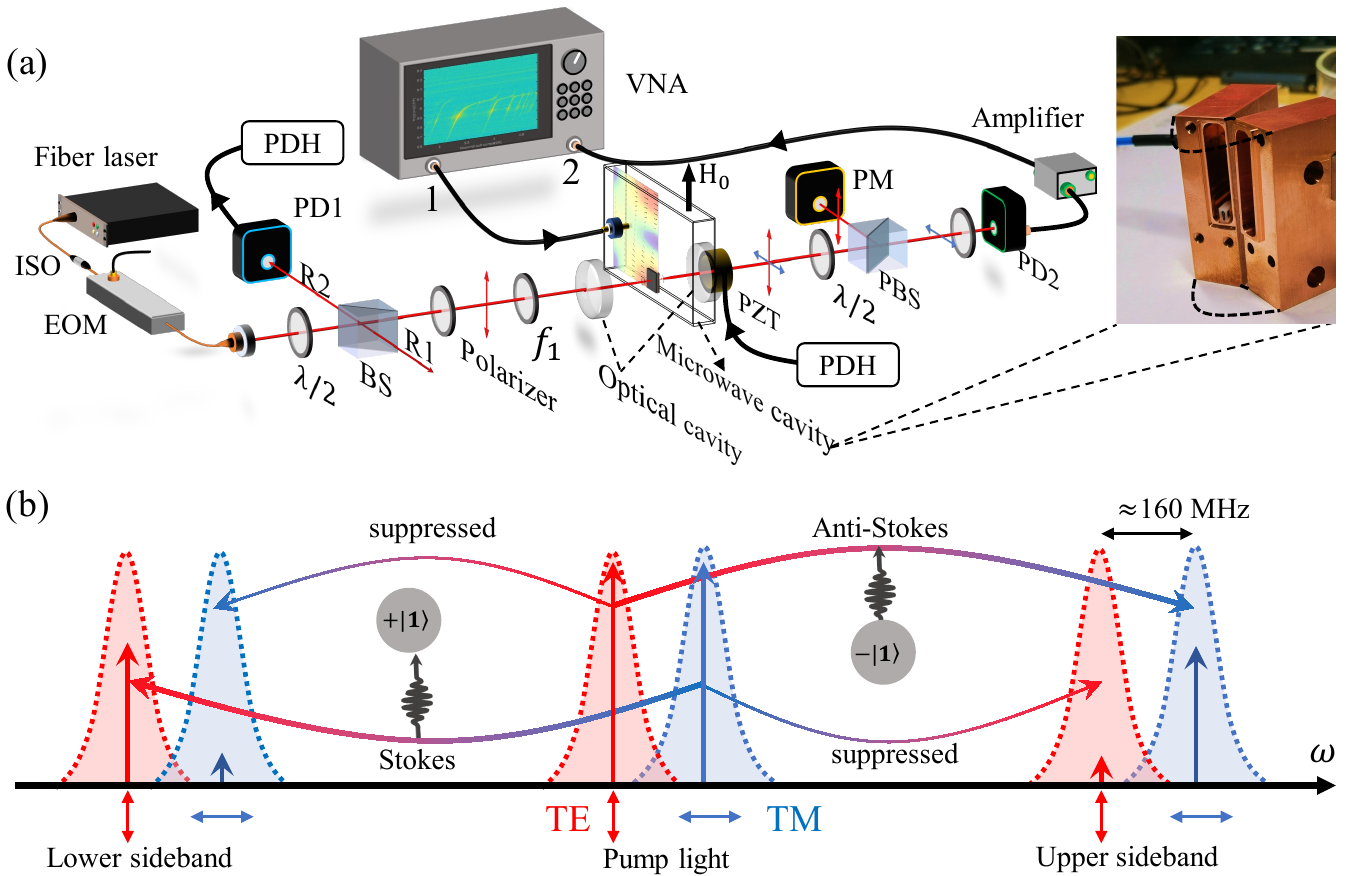}
 \caption{(a) Schematic of the experimental setup. The 1550 nm laser field, modulated by the electro-optical modulator (EOM), is separated by a beam splitter (BS) into the pump light and idler light. The pump light is focused on the YIG sample through lens $f_1$. The polarization of the pump light can be initialized to any angle by the half-wave plate and polarizer. The fast-speed photodetector 1 (PD1) receives the reflected pump field from the optical cavity mirror. The Pound-Drever-Hall (PDH) system controls the EOM, the PD1 and the piezoelectric ceramic transducer (PZT) to lock the optical cavity on the pump light frequency. The output field from the optical cavity is separated into two paths with perpendicular polarizations by the polarization beam splitter (PBS), which enter the power monitor (PM) and the fast-speed photodetector 2 (PD2), respectively. Each beam contains half of the pump light and the converted sideband. The beat signal demodulated by the PD2 is loaded to the vector network analyzer (VNA) after being amplified by the microwave amplifier (AP). The VNA injects microwaves through port 1 into the microwave cavity to excite magnons. (b) The Brillouin scattering selection rule dependent on the polarization of the pump light. The optical cavity has two sets of spectra slightly staggered by about 160 MHz (TE and TM). The pump photon can convert to an optical anti-Stokes (Stokes) sideband photon by absorbing (creating) a magnon.}
	\label{fig1}
\end{figure*}

For the realization of a magnon-based quantum interface in practice, it is vital to strengthen the optomagnonic coupling strength, which is known to be proportional to the square root of the optical photon number ($g_{mb}\propto \sqrt{N_b}$)~\cite{aspelmeyerCavityOptomechanics2014,liuOptomagnonicsMagneticSolids2016}. The experiments have progressed from the single scattering event scheme \cite{hisatomiBidirectionalConversionMicrowave2016a} to the scattering enhancement approach using an optical cavity, where the high-$Q$ cavity greatly increases the number of times the photons are scattered by magnons before they leak out of the cavity. The latter approach has been realized using, e.g., optical whispering gallery mode (WGM) cavities~\cite{chaiSinglesidebandMicrowavetoopticalConversion2022,zhangOptomagnonicWhisperingGallery2016,shenCoherentCouplingPhonons2022}, ridge waveguide cavities~\cite{zhuWaveguideCavityOptomagnonics2020}, and Fabry-P\'erot (FP) nano-cavities~\cite{haighPolarizationDependentScattering2021,pantazopoulosPhotomagnonicNanocavitiesStrong2017}. These cavity architectures can efficiently enhance the mode overlap between the optical cavity mode and the magnon mode. The microwave-optical transduction efficiency can be optimized by leveraging the triple-resonance condition, where the magnon frequency is tuned to match the free spectral range (FSR) of the optical cavity. Using these optical cavity configurations, the current highest conversion efficiency of the microwave-to-optics interface based on magnon modes has reached $\sim 1\times 10^{-8}$~\cite{zhuWaveguideCavityOptomagnonics2020}. However, adjusting the FSR of the optical cavity over a broad range was technically challenging in these previous schemes. In this manner, the advantage of the quantum interface based on the magnon mode, i.e., the tunability of the magnon mode frequency, is actually limited. For a non-adjustable optical cavity with the triple-resonance condition satisfied, the frequency of the magnon mode can only be adjusted to a specific value. Hence, to make full use of the advantages of the magnon modes it is preferable to adopt an optical cavity that has a variable FSR in such a magnon-based microwave-optics interface. On the other hand, there exist multiple magnetostatic modes with non-zero wave vectors in addition to the most commonly studied spin uniform-precession mode, i.e., the Kittel mode, in the YIG samples. Previous studies have not addressed whether these magnetostatic modes have superior effects in the {microwave-optics} conversion.

In this study, an optical cavity that can work at different FSRs is employed. By changing the FSR of the optical cavity and the frequencies of the magnetostatic modes at the same time, the triple-resonance condition and the optimal conditions for the conversion can be easily achieved. By using a YIG flake sample with dimensions of $3\times3\times0.5~\rm{mm}^3$, we make use of its ability to sustain several standing spin wave modes, including magnetostatic surface waves (MSSWs) and  backward volume magnetostatic waves (BVMSWs). Our investigation reveals that the MSSWs demonstrate higher conversion efficiency.  We accordingly achieve an internal conversion efficiency of $1.28\times10^{-7}$, accompanied with a conversion bandwidth of 24 MHz and an adjustable bandwidth of 300 MHz. Nevertheless, we find that the coupling between magnons and optical photons remains the primary obstacle in improving the conversion efficiency, as indicated by our theoretical analysis. Therefore, This finding offers a path to further enhance the conversion efficiency by reducing the mode volume of the optical cavity.

\section{Results and Discussion}
\subsection{Experimental setup and theoretical model}
\begin{figure*}[!t]
	\includegraphics[width=0.9\textwidth]{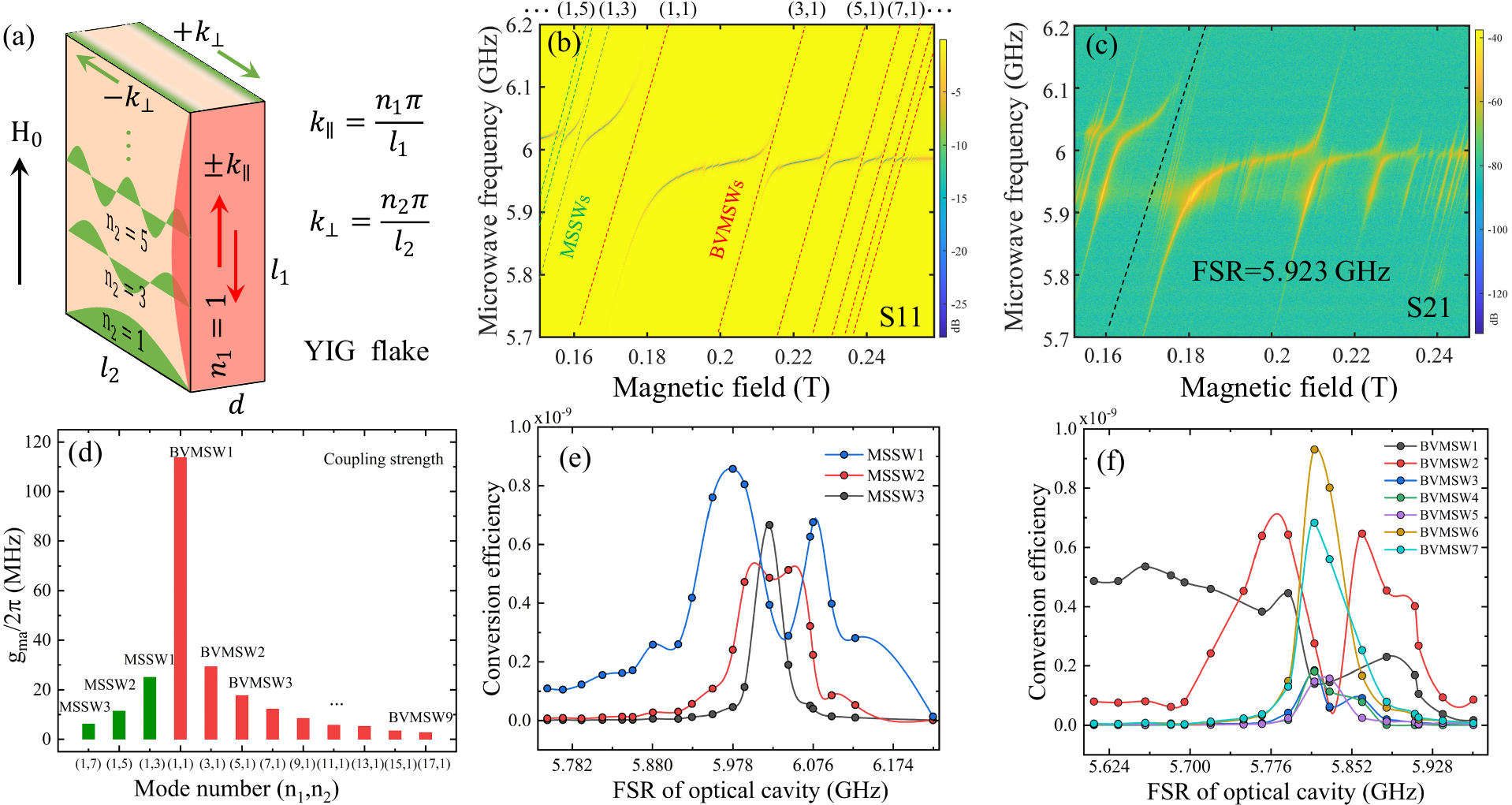}
	\caption{(a) Two main kinds of magnetostatic modes in the in-plane magnetized YIG flake: Bulk modes (BVMSWs) and surface modes (MSSWs) labeled by mode number ($n_1, n_2$). (b) Microwave reflection spectra of the cavity magnonic system versus bias magnetic field. The anti-crossings indicate the strong coupling between the magnetostatic modes and the microwave cavity. The corresponding coupling strengths are  listed in (d). The dispersions of  the MSSWs (green dash lines) and BVMSWs (red dash lines) are plotted by using Eqs.~(\ref{MSSW}) and (\ref{BVMSW}), respectively. (c) The S21 mapping corresponds to the beat signal of the pump field and the converted sideband, which is demodulated by the PD2 and measured by the VNA. The brightest spot corresponds to the optical cavity enhanced microwave-to-optics conversion, where the triple-resonance condition is satisfied. The corresponding FSR of the optical cavity is 5.923 GHz and the incident optical power is 9.3 dBm. (e), (f) The photon-number conversion efficiency of the MSSWs and BVMSWs at various FSRs.}
	\label{fig2}
\end{figure*}

The schematic of the experimental setup is illustrated in Fig.~\ref{fig1}(a). The microwave-optomagnonic system contains a microwave cavity, a YIG flake sample, and an optical FP cavity. We employ two kinds of microwave cavities in the experiment, which are respectively a 3D microwave cavity and a 2D split-ring resonator (SRR). The details of the 3D cavity are given in the main text and those of the SRR are presented in Appendix A. The 3D cavity is made of the oxygen-free copper with inner dimensions of 28$\times$56$\times$6 mm$^3$ and the eigenfrequency of the $\rm{TE_{101}}$ mode is 5.99 GHz. The YIG flake sample, which is a single crystal grown along the $\langle 111\rangle$ axis, has been coated with anti-reflective optical coatings to minimize the optical loss. The YIG sample is positioned at the antinode of the magnetic field of the cavity $\rm{TE_{101}}$ mode, and a static magnetic field $H_0$ is uniformly applied to the whole sample.

As shown in Fig.~\ref{fig1}(a), the light field passes through the microwave cavity and the YIG sample via the holes on the walls of the microwave cavity. The optical cavity is formed by two axially movable optical mirrors. The port~1 of the vector network analyzer (VNA) is connected to the microwave cavity to excite the magnon modes in the YIG sample and the port 2 is used to receive the beat signal of the {pump light} and the optical sidebands scattered by magnons, which include both the Stokes and anti-Stokes sidebands. The 1550 nm laser field is separated by the beam splitter (BS) into two paths: the pump field, also known as the carrier field, and the reflected field (R1). The pump field is loaded to the optical cavity and R1 is  put on standby. The second reflected light field (R2) originates from the pump field that is reflected by the optical cavity mirror. It is employed to stabilize the optical cavity at the frequency of the pump field using the Pound-Drever-Hall (PDH) control system. The details of the PDH control system are provided in Appendix A.

After passing through the optical cavity, the optical beam, which comprises the residual pump field and the scattered optical sideband with an orthogonal polarization, encounters a half-wave plate and is subsequently separated by the polarization beam splitter (PBS). Then, the separated beams are respectively loaded to the power meter (PM) and the fast photodetector (PD2). Each beam contains an equal portion of the residual pump field and the sideband signal. The beat signal generated by the pump field and the sideband at PD2 is amplified by the microwave amplifier before being transferred to the VNA. Meanwhile, the PM monitors the output light power in time.  One can confirm that the polarizations of the optical sideband and the pump field are perpendicular to each other by two facts: (i) there is no beat signal at the PD2 without the PBS, and (ii) the beat signal appears with a periodicity of $\pi/4 $ by rotating the half-wave plate before the PBS.

To achieve a stronger sideband signal, it is essential to ensure that the optical cavity operates in a state of both mode matching and impedance matching, and satisfies the triple-resonance condition. These requirements can be attained by the following methods. To begin with, the incident laser field is adjusted to match the fundamental Gaussian mode of the optical cavity by tuning the lens $f_1$. Furthermore, impedance matching often necessitates that the reflectivity of the incident mirror be lower than that of the output mirror, which are respectively $99\%$ and $99.5\%$ in our configuration. This is done to achieve a balance between the external coupling strength and the overall attenuation rate of the optical cavity. The outer surfaces of the optical mirrors and the flake sample are covered with an antireflection layer that has a reflectance of less than $0.1\%$. Moreover, the triple-resonance condition can be achieved by adjusting the magnon frequency such that the scattered sideband  photons and the pump field are both in resonance with the optical cavity.

The schematic of the optical cavity modes are illustrated in Fig.~\ref{fig1}(b). Due to the birefringence effect of the YIG material, there are two distinguishable sets of optical resonant eigenmodes, which are represented by the red and blue dashed Lorentz curves. These modes have a frequency difference of around 160 MHz and are polarized perpendicularly. The FSR of the optical cavity can be adjusted by {changing the cavity length}. The parameters of the optical cavity are as follows: the external coupling strength $\kappa_b/2\pi = 6.56$ MHz, the intrinsic dissipation rate $\gamma_b/2\pi =25.14 $ MHz, finesse $F = 173.3$, and the quality factor $Q=6.09\times 10^6$.

When the pump light interacts with the magnon mode, Brillouin light scattering takes place, producing two sidebands in the output field of the optical cavity, i.e., the lower and upper sidebands in Fig.~\ref{fig1}(b), corresponding to the Stokes and anti-Stokes processes, respectively. The total Hamiltonian of the system in the frame rotating at the pump frequency  $\omega_{p}$ is given by (see Appendix B):
\begin{equation}
\begin{split}
	\label{totalH}
	H/\hbar = &\omega_{a} a^{\dagger} a+\omega_{m} m^{\dagger} m+\Delta_{b} b^{\dagger} b
		 + g_{ma} \left( ma^{\dagger} +m^{\dagger}a \right) \\
	 	&+ g_{mb} \left( mb^{\dagger} +m^{\dagger}b \right)
	 	+ g_{mb} \left( mb +m^{\dagger}b^{\dagger} \right),
\end{split}
\end{equation}
where $a~(a^\dagger)$, $m~(m^\dagger)$, and $b~(b^\dagger)$ are the annihilation (creation) operators of the microwave cavity mode, magnon mode, and scattered optical sideband mode, respectively. $\omega_{i},~i=a,m,b$, are the corresponding  resonance frequencies of the modes. $\Delta_{b} =\omega_{b}- \omega_{p}$, and $g_{ma}$ ($g_{mb}$) is the coupling strength between the magnon mode and the microwave (optical) cavity.  Note that $g_{mb}$ is the effective optomagnonic coupling strength enhanced by the strong pump light.
The Stokes process corresponds to a pump photon converting to a lower-frequency sideband photon of frequency $\omega_{p}-\omega_{m}$ ($\Delta_b=-\omega_m$) and simultaneously creating a magnon of frequency $\omega_m$. In contrast, the anti-Stokes process involves the concurrent annihilation of a pump photon and a magnon,  producing a sideband photon of frequency $\omega_{p}+\omega_{m}$ ($\Delta_b=\omega_m$). The polarization of the incident pump light affects the relative intensity of the above two processes. As illustrated in Fig.~\ref{fig1}(b), the scattering probability of the Stokes process is much higher when the pump light is polarized in the transverse magnetic (TM) direction, which is parallel to the external magnetic field. In this case, the anti-Stokes process is strongly suppressed, exhibiting a suppression ratio of 25~dB in our experiment. In contrast, the anti-Stokes scattering becomes dominant when the pump light is in transverse electric (TE) polarization. This scattering selection rule has also been observed in other systems \cite{haighPolarizationDependentScattering2021,prabhakarEffectsHighMicrowave1996}. We can now define the conversion efficiency in the microwave-magnon-light conversion process introduced above. The photon-number conversion efficiency is defined as the ratio of the number of converted optical photons from the output field of the optical cavity to that of the microwave photons from the input field of the microwave cavity. The conversion efficiency $\eta_{as}$ ($\eta_{s}$) in the anti-Stokes (Stokes) process is given by
\begin{equation}
	\label{converrsionefficiency0}
	\eta_{as(s)}=\left | \frac{g_{ma}g_{mb}\sqrt{\kappa _a \kappa _b} }{\chi _a^{-1}\chi _b^{-1}\chi _m^{-1}+g_{ma}^2\chi _b^{-1}\pm g_{mb}^2\chi _a^{-1}}  \right |^2,
\end{equation}
where $\chi_i~(i=a,m,b)$ are the susceptibilities of the microwave cavity, magnon mode, and optical cavity, respectively, of which the expressions are provided in Appendix B. $\kappa_{a(b)}$ is the external coupling rate of the microwave (optical) cavity. The anti-Stokes and Stokes conversion efficiencies are essentially equivalent for a moderate coupling strength $g_{mb}$ and become obviously different when $g_{mb}$ is strong. This point is further expanded in Appendix B.

\begin{figure*}[!t]
	\includegraphics[width=0.95\textwidth]{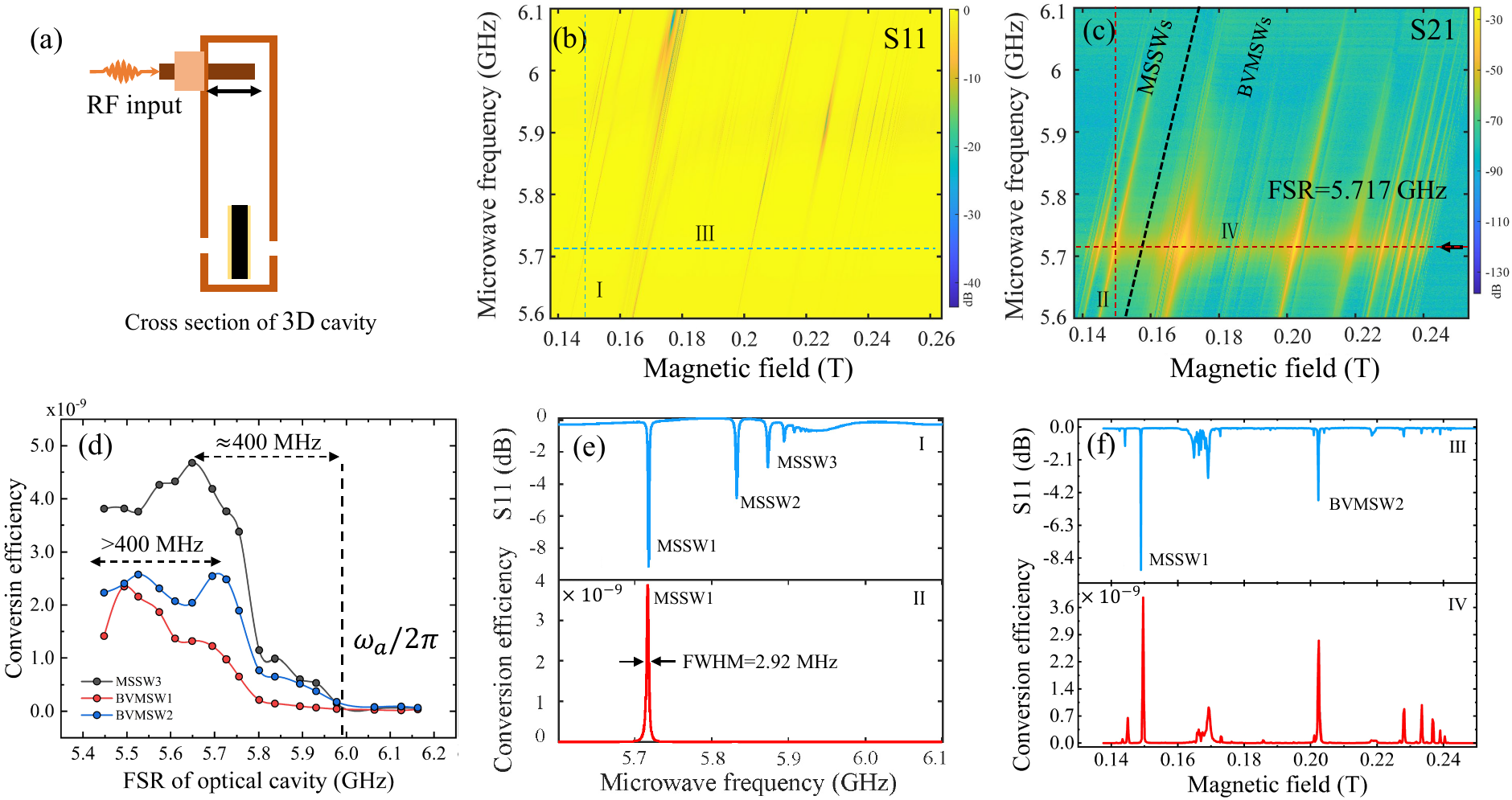}
	\caption{  (a) The length of the pin into the cavity can be adjusted to change the external coupling rate of the microwave cavity.   (b) Microwave reflection spectra of the cavity magnonic system  versus the bias magnetic field for the increased external coupling of the microwave cavity. (c) The microwave-to-optical photon conversion spectra. The FSR of the optical cavity is adjusted to 5.717 GHz. (d) The photon-number conversion efficiency at various FSRs. The optimal conversion spot is about $400$ MHz away from the cavity mode and the tunable bandwidth is around 400 MHz.  (e) Reflection (microwave-to-optics conversion efficiency) spectrum extracted from (b) [(c)] at the magnetic field corresponding to the vertical line \uppercase\expandafter{\romannumeral1} (\uppercase\expandafter{\romannumeral2}). (f) Reflection (conversion efficiency) spectrum extracted from (b) [(c)] at the fixed microwave frequency indicated by the horizontal line \uppercase\expandafter{\romannumeral3} (\uppercase\expandafter{\romannumeral4})}.
	\label{fig3}
\end{figure*}

\subsection{Microwave-optical photon conversion in the 3D microwave cavity configuration}

In the finite-size YIG flake, magnetostatic modes are characterized as distinct standing wave modes based on the spin wave vector and spatial distribution. When using an in-plane magnetized YIG flake, as depicted in Fig.~\ref{fig2}(a), the parallel and vertical components of the spin wave vector are defined as $k_{\parallel}=n_1 \pi/ l_{1}$ and $k_{\perp}=n_2 \pi/ l_{2} $, respectively~\cite{storeyDeterminationFrequenciesMagnetostatic,pirroAdvancesCoherentMagnonics2021}, where $n_1$ and $n_2$ are positive integers, $l_{1}$ and $l_{2}$ are the side lengths of the sample. Correspondingly, a mode number pair of $(n_1,n_2)$ can specifically define a spin wave mode in the finite-size flake sample. In the experiment, we focus on two main types of magnonic standing wave modes: Magnetostatic surface waves (MSSWs) with mode number (1,$n_2$) and backward volume magnetostatic waves (BVMSWs) with mode number ($n_1$,1). The MSSWs are primarily located near the surface of the flake, while the BVMSWs get their name because their distribution extends throughout the bulk of the flake. Additionally, the phase and group velocities of the BVMSWs within the flake proceed in opposite directions~\cite{SpinWavesTheory2009a}. The dispersion relations of the MSSWs and BVMSWs are as follows:
\begin{subequations}
	\begin{align}
		\label{MSSW}
		\omega_{\rm{S}}^{2} &=\omega_{0}\left(\omega_{0}+\omega_{\rm{M}}\right)+\frac{\omega_{\rm{M}}^2 }{4} \left[ 1-e^{-2kd}  \right],\\
		\label{BVMSW}
    	\omega_{\rm{B}}^{2} &=\omega_{0}\left[\omega_{0}+\omega_{\rm{M}}\left(\frac{1-e^{-k d}}{k d}\right)\right],
	\end{align}
     \label{dispersionrelationship}
\end{subequations}
where $\omega_0$=$-\gamma\mu_0H_0$, $\omega_{\rm{M}}$=$-\gamma \mu_0 H\rm{_M}$, $H\rm{_M} $ is the saturation magnetization of the YIG, $k$ is the spin wave vector, and $d$ is the thickness of the flake sample. $\gamma/2\pi$ $\approx$ $28~ \rm{GHz/T}$ is the gyromagnetic ratio and $\mu_0 $ is the vacuum permeability. In Fig.~\ref{fig2}(b), the MSSWs and BVMSWs are represented by the green and red dashed lines, respectively. These lines are plotted using Eq.~(\ref{dispersionrelationship}). The strong interaction between the microwave cavity mode and magnetostatic mode is evident in the repulsion of energy levels. The coupling strengths are depicted in Fig.~\ref{fig2}(d).

Following that, the optical pump is turned on and the FSR of the optical cavity is set to 5.923 GHz. The conversion of microwave to optical photons is measured by gradually changing the probe frequency and magnon mode frequency, as shown in Fig.~\ref{fig2}(c). The contour plot illustrates the efficacy of converting microwave to light waves in various regions. It has been observed that cavity-magnon polaritons, not the pure magnon modes, serve as the medium of conversion. Enhanced conversion occurs in luminous regions where the triple-resonance condition is satisfied.

We further tune the FSR of the optical cavity to make the cavity-enhanced conversion occur at various polariton frequencies. The microwave-optical photon conversion efficiencies via the MSSW modes and BVMSW modes are calculated using Eq.~(\ref{converrsionefficiency0}) and the results are shown in Fig.~\ref{fig2}(e) and Fig.~\ref{fig2}(f), respectively. The results indicate that there is no significant difference in the maximum conversion efficiency ($9.31 \times 10^{-10}$) between the MSSW modes and BVMSW modes in the 3D microwave cavity configuration. However, spin wave modes that have a stronger interaction with the microwave cavity exhibit wider conversion bandwidths. For instance, the BVMSW1 has an effective conversion range of up to 200 MHz.

 We then push the microwave input-output pin deeper into the microwave cavity, resulting in an increased external coupling rate of the microwave cavity mode. The microwave reflection and conversion mappings with respect to the bias magnetic field are shown in Fig.~\ref{fig3}(b)-\ref{fig3}(c). The increased external coupling rate straightens the dispersion of polaritons. With an increased external coupling rate, which yields a higher pump efficiency, more microwave photons can enter the microwave cavity (for a fixed pump field) to interact with the magnons. Consequently, we achieve enhanced conversion efficiencies, which are also evaluated at various FSR values, as illustrated in Fig.~\ref{fig3}(d). The typical bandwidth exceeds 400 MHz, and the optimal conversion point is significantly distant from the microwave cavity resonance frequency. As shown in Fig.~\ref{fig3}(e)-\ref{fig3}(f), the MSSWs family exhibits both higher conversion efficiency and greater mode resolution compared to the BVMSWs family. This is because the BVMSWs have a larger mode volume, which is more easily affected by the inhomogeneity of the microwave field. It is evident that the advantage of the MSSWs may persist even in a planar cavity with a higher level of non-uniformity of the electromagnetic field, as illustrated in the following section.

\begin{figure*}[t]
	\includegraphics[width=0.95\textwidth]{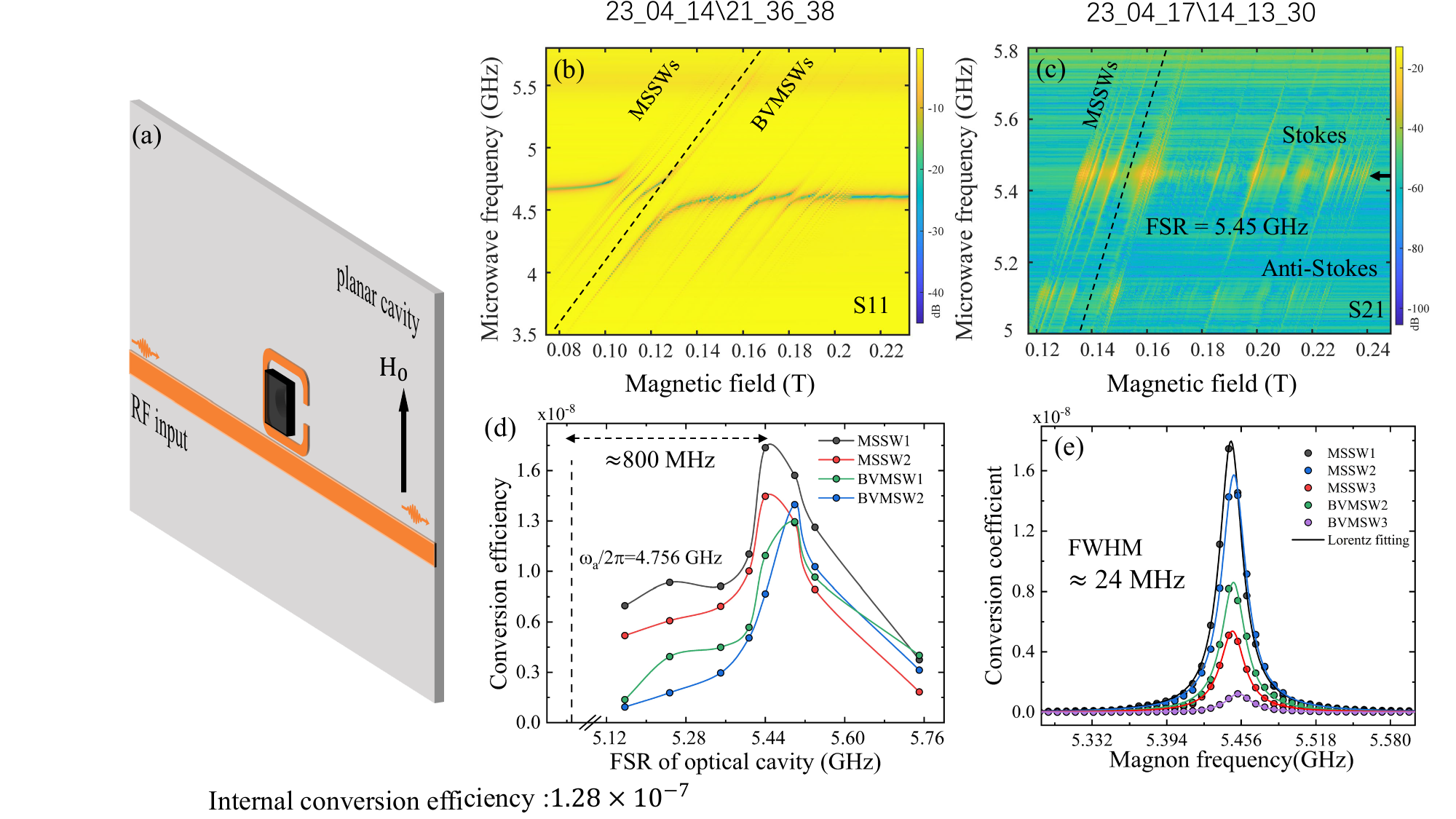}
	\caption{(a) A planar cavity made of a split-ring resonator on the printed circuit board couples with the magnon modes of the YIG sample placed near the ring center. (b) Microwave reflection spectra of the cavity magnonic system. (c) The microwave-to-optical photon conversion spectra. The Stokes (anti-Stokes) sidebands are enhanced (suppressed) for the TM polarized pump light, cf. Fig.\ref{fig1}(b). (d) The conversion efficiencies associated with the Stokes sidebands versus the FSR of the optical cavity. (e) The Lorentz function is used to fit the conversion bandwidths of the magnetostatic modes, with full width at half maximum about 24 MHz.}
	\label{fig4}
\end{figure*}

\subsection{Microwave-optical photon conversion in the planar microwave cavity configuration}

To further improve the conversion efficiency, we employ a split-ring resonator (SRR) cavity to interact with the YIG sample, as illustrated in Fig.~\ref{fig4}(a). The planar cavity configuration exhibits a higher degree of localization of the electromagnetic field, leading to a stronger coupling between microwaves and magnons. The measurement setup remains the same as in Fig.~\ref{fig1}(a). The resonance frequency of the SRR cavity is specifically set to 4.6 GHz. The reflection spectra versus the microwave frequency measured at different bias magnetic  fields are displayed in Fig.~\ref{fig4}(b). In comparison to the 3D cavity, the planar device excites a number of higher-order magnetostatic modes, particularly the bulk modes (BVMSWs). This is due to the increased inhomogeneity of the microwave magnetic field in the planar device. In the measurement of the microwave-optics conversion, we first set the FSR of the optical cavity to 5.45 GHz. We then proceed to detect both the Stokes and anti-Stokes sidebands, as shown in Fig.~\ref{fig4}(c). The suppression ratio of the enhanced Stokes sideband to the suppressed anti-Stokes sideband reaches 25 dB. We regularly modify the FSR to explore the relation between the conversion efficiency and the FSR, and find that the frequency at which the conversion efficiency is highest is far from the resonance frequency of the microwave cavity, with the discrepancy of about 800 MHz as indicated in Fig.~\ref{fig4}(d). Accordingly, we attain a photon-number conversion efficiency of $\eta = 1.75 \times 10^{-8}$ for the MSSW1 in the measurement, which has surpassed the previous record ($1.08 \times 10^{-8}$)~\cite{zhuWaveguideCavityOptomagnonics2020}. The corresponding internal conversion efficiency is $1.28\times10^{-7}$. Meanwhile, it can be observed that surface spin wave modes contribute to a higher conversion efficiency. Furthermore, in accordance with the 3 dB concept, the adjustable range of the microwave-optical conversion bandwidth surpasses 300 MHz. The conversion bandwidth for each individual magnetostatic mode is approximately 24 MHz, as shown in Fig.~\ref{fig4}(e).

\section{Conclusion}
Achieving high-efficiency conversion of microwaves to optical signals is essential for the successful implementation of quantum interfaces. The magnonic system has emerged as a potential platform for this goal, thanks to its versatile adaptability and capacity to interact with diverse quantum systems. Nevertheless, the low conversion efficiency of the magnonic system is attributed to the weak optomagnonic coupling strength. To address this issue, the present work employs two kinds of microwave cavities and an adjustable high-$Q$ optical cavity to enhance the microwave-magnon-light conversion efficiency. By adjusting the microwave input-output pin of the 3D cavity or using a planar cavity, the conversion efficiency is prominently improved, at the price of stimulating more higher-order bulk modes. By achieving the triple-resonance condition,  we attain a highest photon-number conversion efficiency of $1.75\times10^{-8}$, corresponding to an internal conversion efficiency of $1.28\times10^{-7}$. The conversion process operates within a typical single-mode bandwidth of 24 MHz, and the conversion bandwidth could be adjusted up to 300 MHz. In addition, we find that the MSSW modes exhibit higher conversion efficiency compared to the BVMSW modes, mostly due to their higher mode density. The current design suffers from the small overlap between the YIG sample and the optical cavity, which limits the conversion efficiency. A remarkable improvement would be achieved by significantly reducing the mode volume of the optical cavity. Our present work opens an avenue towards further enhancing and facilitating the microwave-to-optical photon conversion based on optomagnonic systems.


\appendix
\renewcommand{\appendixname}{APPENDIX}
\section{Experimental details}

\noindent\textbf{Light source and detection devices}

\noindent The laser field is generated by the NKT Koheras BASIK E15 1550~nm fiber laser. The laser power is adjusted from 12 mW to 40 mW, but remains from 2 mW to 8 mW when reaching the optical cavity due to the loss caused by the BS and EOM (Thorlab LNP6119). The biased detector PD1 (Thorlab DET10N2) is used for the PDH system. The detector PD2 is a fiber fast-speed detector (NEW Focus 818-BB-35F) with responsivity 0.68 A/W at 1550 nm, which generates the beat signal between the sidebands and the laser source field. The beat  (microwave) signal is amplified by the low-noise amplifier (Mini-Circuits 2x60-83LN) for 39.5 dB. The power monitor used is Thorlab PM103 S122C. The microwave signals are analyzed by vector network analyzer (Keysight ENA5071C).
\\

\noindent\textbf{Design of the YIG sample and optical cavity}

\noindent The YIG slice of $3\times3\times0.5~\rm{mm^3}$ is cut from bulk materials growth along the  $\langle111\rangle$ crystal axis. The loss coefficient of the coated YIG sample is about 5485 ppm. The intrinsic optical cavity loss coefficient is about 4500 ppm, which is mainly induced by the mirror loss and imperfect mode matching. To reduce the optical loss, the inner surfaces of the cavity mirrors are coated with high-reflection films: Specifically, the left one has a reflection of $R_{L}= 99\% $ and the right one is $R_{R}=99.5\%$. In addition, both sides of the YIG sample and the outer surfaces of the cavity mirrors are coated with the anti-reflection dielectric films ($R<0.1\%$) to reduce the scattering loss. In this case, the external coupling and intrinsic dissipation of the optical cavity with sample are $\kappa_b/2\pi = 6.56$ MHz and $\gamma_b/2\pi =25.14 $ MHz, respectively.
\\

\noindent\textbf{Design of the planar microwave cavity}

\begin{figure*}[!t]
	\includegraphics[width=0.95\textwidth]{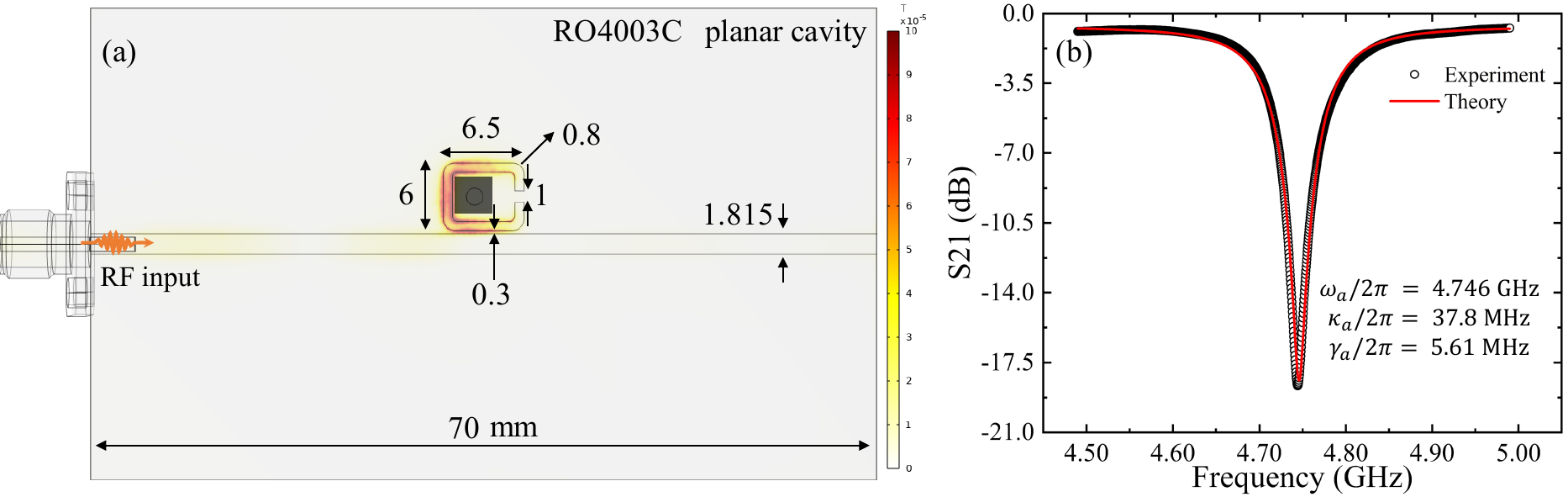}
	\caption{(a) The parameters of the planar cavity and the distribution of the microwave magnetic field at the cavity resonance frequency. The YIG flake is placed near the center of the split-ring resonator. (b) The resonance frequency and dissipation rates of the planar cavity are obtained by fitting its microwave reflection spectrum.}
	\label{Sfig1}
\end{figure*}

\noindent  The planar split-ring cavity is fabricated using the Rogers RO4003C board, which consists of two layers of 35 $\mu$m thick copper sandwiched by a 0.8 mm dielectric.  The split-ring resonator is side-coupled to the transmission line, as shown in Fig.~\ref{Sfig1}(a). The device is processed by a circuit board engraving machine (LPKF ProtoMat S104). The colorbar shows the distribution of the microwave magnetic field at the cavity resonance frequency. The YIG slice (grey area) is placed near the center of the ring. The mode density of the planar cavity is higher compared with the 3D cavity. The resonance frequency and the dissipation rates are obtained by fitting the transmission spectrum, which is shown in Fig.~\ref{Sfig1}(b).
\\

\noindent\textbf{PDH locking cavity system }

\noindent The PDH (Pound-Drever-Hall) system \cite{dreverLaserPhaseFrequency1983} is used for locking cavity resonance or stabilizing the laser frequency. It can lock the optical cavity on the pump light frequency, in spite of the presence of the low-frequency mechanical vibration and frequency fluctuation of the pump light. The PDH is essentially a negative feedback system. In our case, it works in the following way: The power of the light reflected by the cavity mirror is monitored and maintained at the minimum value, corresponding to the cavity resonance, by providing feedback voltage to the PZT to change the optical cavity length in real time.

\section{Theory and numerical analysis}

\begin{figure*}[!t]
	\includegraphics[width=0.95\textwidth]{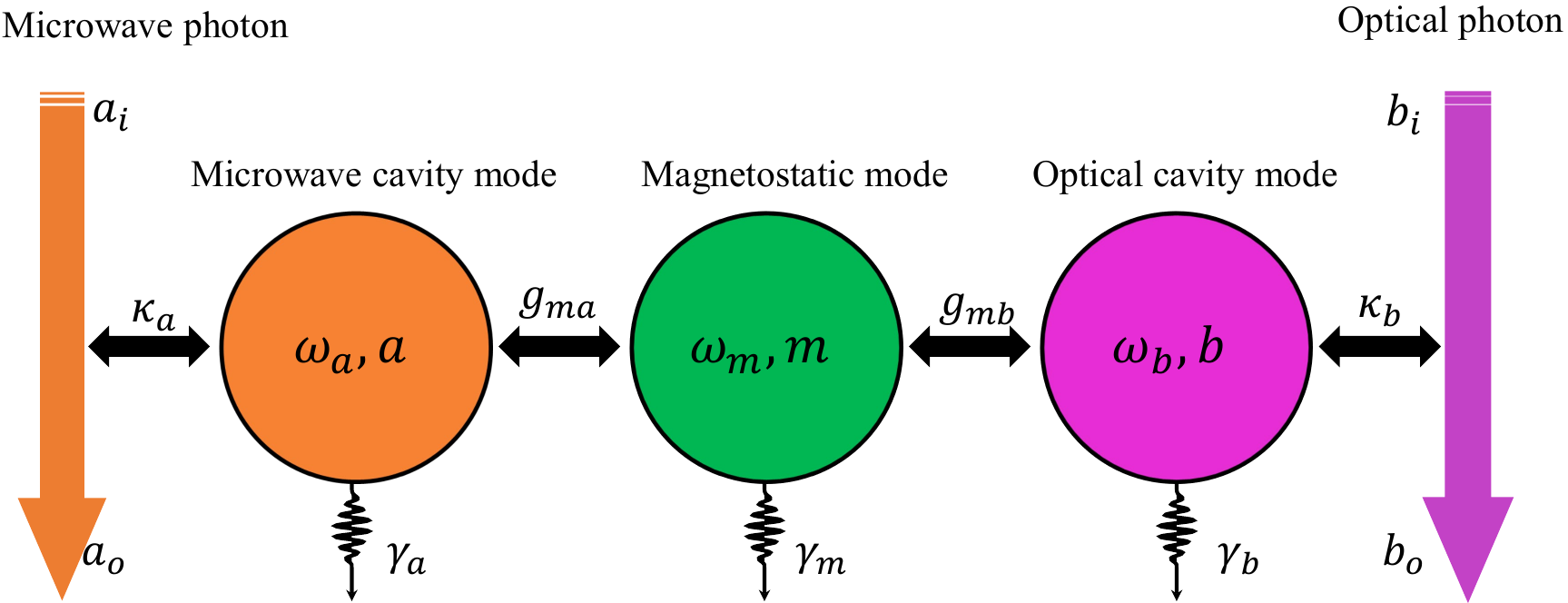}
	\caption{ The diagram of the microwave-to-optics conversion. The microwave cavity is coherently coupled with the magnon mode with the coupling strength $g_{ma}$, and the magnon mode is also coupled with the optical cavity with the coupling strength $g_{mb}$. The external coupling rate of the microwave (optical) cavity is $\kappa_a$ ($\kappa_b$). The intrinsic dissipation rates of the three modes are $\gamma_{a,m,b}$, and $a_{i,o}$ ($b_{i,o}$) denote the input and output fields of the microwave (optical) cavity. }
	\label{Sfig2}
\end{figure*}

The system under consideration comprises a microwave cavity mode, a magnon mode, and two optical modes, as modeled in Fig.~\ref{Sfig2}. The total Hamiltonian of the system is given by
\begin{equation}
	\label{A0}
	H = H_0 + H_I.
\end{equation}
 The first term is the free Hamiltonian of the system, i.e.
\begin{equation}
	\label{A1}
	H_0 /\hbar= \omega_a a^{\dagger}a + \omega_m m^\dagger m + \omega_b b^\dagger b + \omega_p p^\dagger p ,
\end{equation}
where $a~(a^\dagger),~m~(m^\dagger),~b~(b^\dagger),~p~(p^\dagger)$ are the annihilation (creation) operators of the microwave cavity mode, the magnon mode, the converted optical sideband mode, and the optical pump mode, respectively. $\omega_{i}$, $i=a,m,b,p$, are the corresponding resonance frequencies. The second term describes the interactions among the four modes, which is given by
\begin{equation}
	\label{A3}
	H_I/\hbar =  g_{ma} \left ( ma^{\dagger} +m^{\dagger}a\right ) + g'_{mb} \left ( m +m^{\dagger}\right )\left ( pb^{\dagger}+ p^{\dagger}b\right ) ,
\end{equation}
where $g_{ma}$ is the coherent coupling strength between the magnon mode and the microwave cavity mode. $g'_{mb}$ represents the single-photon optomagnonic coupling strength. Since the pump light is intense, it can be treated classically by replacing the operators with complex numbers, i.e., $p =\beta$, $p^{\dagger} =\beta^*$. For simplicity, we choose the phase reference, such that $\beta$ is real and positive, i.e., $\beta^*=\beta$.  Then, the interaction Hamiltonian $H_I$ leads to a simplified linear form and can be decomposed into the ones for describing, respectively, the anti-Stokes and Stokes processes, which are
\begin{subequations}
	\begin{align}
		\label{A3}
		H_{I}^{as}/\hbar =  g_{mb} \left ( mb^{\dagger} +m^{\dagger}b\right )  ,\\
		H_{I}^{s}/\hbar =  g_{mb} \left ( mb +m^{\dagger}b^{\dagger}\right ) ,
		\label{heat}
	\end{align}
\end{subequations}
where $g_{mb}= g'_{mb} \beta$ is the pump-enhanced effective optomagnonic coupling strength. In what follows, we discuss the anti-Stokes and Stokes processes separately.

 In the frame rotating at the pump frequency $\omega_p$, we obtain the following Langevin equations of the system for the anti-Stokes process:
 \begin{subequations}
	\begin{align}
	\dot a &=  - i{\omega _a}a - \frac{{{\kappa _a} + {\gamma _a}}}{2}a - i{g_{ma}}m + \sqrt {{\kappa _a}} {a_i}\\
	\dot m &=  - i{\omega _m}m - \frac{{{\gamma _m}}}{2}m - i{g_{ma}}a - i{g_{mb}}b\\
	\dot b &=  - i{\Delta _b}b - \frac{{{\kappa _b} + {\gamma _b}}}{2}b - i{g_{mb}}m + \sqrt {{\kappa _b}} {b_i}
	\end{align}
\end{subequations}
where $\Delta_{b} =\omega_{b}- \omega_{p}$, $\kappa_{a}$ ($\kappa_{b}$) is the external coupling rate of the microwave (optical) cavity, $\gamma_{a}$ ($\gamma_{b}$) is the intrinsic dissipation rate of the microwave (optical) cavity, and  $\gamma_{m}$ is the dissipation rate of the magnon mode. The corresponding Langevin equations for the Stokes process are given by
\begin{subequations}
	\begin{align}
		\dot a &=  - i{\omega _a}a - \frac{{{\kappa _a} + {\gamma _a}}}{2}a - i{g_{ma}}m + \sqrt {{\kappa _a}} {a_i}\\
		\dot m &=  - i{\omega _m}m - \frac{{{\gamma _m}}}{2}m - i{g_{ma}}a - i{g_{mb}}{b^\dag }\\
		{{\dot b}^\dag } &= i{\Delta _b}b^{\dag} - \frac{{{\kappa _b} + {\gamma _b}}}{2}{b^\dag } + i{g_{mb}}m + \sqrt {{\kappa _b}} {b_i}^\dag
	\end{align}
\end{subequations}
Taking the Fourier transform and solving the above two sets of equations in the frequency domain, we obtain the steady-state solutions of the system, which are
\begin{subequations}\label{antiStoEq}
	\begin{align}
		\label{QLEa}
		a &=\chi _{a}\left [ -ig_{ma} m+\sqrt{\kappa _{a} }a_{i}     \right ],  \\
		\label{QLEb}
		m &=\chi _{m}\left [ -ig_{ma} a- ig_{mb} b   \right ],  \\
		\label{QLEd}
		b &=\chi _{b}\left [ -ig_{mb} m+\sqrt{\kappa _{b} }b_{i}     \right ],
	\end{align}
\end{subequations}
for the anti-Stokes process, and
\begin{subequations}\label{StoEq}
	\begin{align}
		\label{QLEa}
		a &=\chi _{a}\left [ -ig_{ma} m+\sqrt{\kappa _{a} }a_{i}     \right ],  \\
		\label{QLEc}
		m &=\chi _{m}\left [ -ig_{ma} a- ig_{mb} b^{\dagger }     \right ],  \\
		\label{QLEe}
		b^{\dagger} &=\chi _{b}\left [ ig_{mb} m+\sqrt{\kappa _{b} }b_{i}^{\dagger }     \right ],
	\end{align}
\end{subequations}
for the Stokes process. The susceptibilities $\chi_{i}$ are defined as
\begin{subequations}
	\begin{align}
		\chi _{a} &=\left [ -i\left ( \omega -\omega _{a} \right ) +\frac{\gamma _{a} +\kappa _{a} }{2}   \right ]^{-1},  \\
		\chi _{m} &=\left [ -i\left ( \omega -\omega _{m} \right ) +\frac{\gamma _{m}  }{2}   \right ]^{-1},  \\
		\label{xic}	
		\chi _{b} &=\left [ -i\left ( \omega -\omega _{m} \right ) +\frac{\gamma _{b} +\kappa _{b} }{2}   \right ]^{-1}.
	\end{align}
\end{subequations}
Note that the expression of the susceptibility $\chi_{b}$ is provided when the triple-resonance condition is satisfied, i.e., $\Delta_b=\omega_m$ for the anti-Stokes process and $\Delta_b=-\omega_m$ for the Stokes process.

By fully solving Eqs. \eqref{antiStoEq} and \eqref{StoEq}, and utilizing the input-output relations $b_{o}= \sqrt{\kappa_b }b -b_{i} $ and $b_{o}^{\dagger}= \sqrt{\kappa_b } b^{\dagger} -b_{i}^{\dagger}$, we can obtain the output field of the optical cavity, which allows us to define the microwave-to-optics conversion efficiency in the anti-Stokes or Stokes process, i.e.,
\begin{subequations}
	\label{conversionefficiency}
	\begin{align}
		\label{conversionefficiencya}
		\eta_{as} &=\left |  \frac{b_{o}}{a_i} \right |^2 \nonumber\\
		&=\left | \frac{g_{ma}g_{mb}\sqrt{\kappa _a \kappa _b} }{\chi _a^{-1}\chi _b^{-1}\chi _m^{-1}+g_{ma}^2\chi _b^{-1}+g_{mb}^2\chi _a^{-1}}  \right |^2, \\
		\eta_{s} &=\left |  \frac{b_{o}^{\dagger}}{a_i} \right |^2 \nonumber\\
		&=\left | \frac{-g_{ma}g_{mb}\sqrt{\kappa _a \kappa _b} }{\chi _a^{-1}\chi _b^{-1}\chi _m^{-1}+g_{ma}^2\chi _b^{-1}-g_{mb}^2\chi _a^{-1}}  \right |^2.
		\label{conversionefficiencys}
	\end{align}
\end{subequations}
It can be seen that the conversion efficiencies $\eta_{\rm{as}} $ and $\eta_{s}$ in the two processes are nearly identical for a small optomagnonic coupling strength $g_{mb}$. We further define the internal conversion efficiency $\eta_{\rm{int}}$ by factoring out the coupling losses from the total conversion efficiency, i.e., $\eta_{\rm{int}}=  \eta_{as(s)}/(\xi _{a} \xi  _{b})$, where $\xi _{i}=\kappa_i/(\kappa_i+\gamma_i) $ for $i = a,b$.

\begin{figure*}[!t]
	\includegraphics[width=0.95\textwidth]{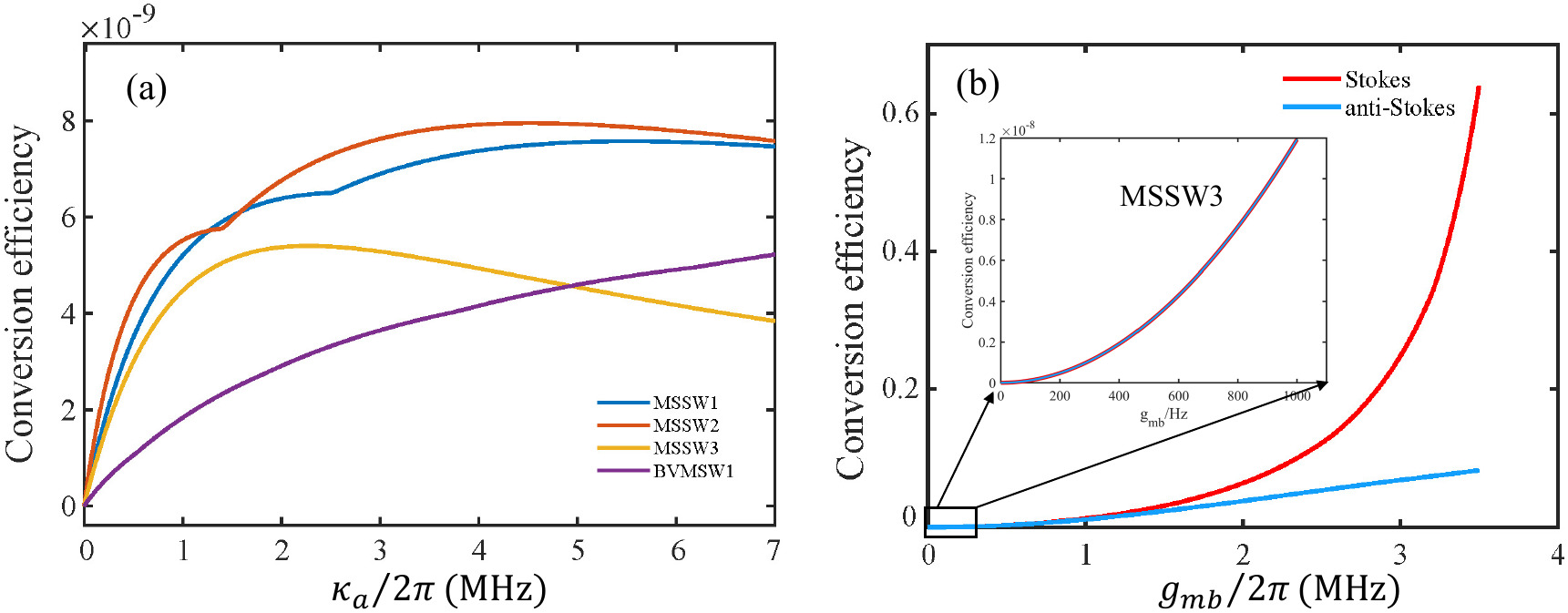}
	\caption{(a) The conversion efficiency versus the external coupling rate $\kappa_a$ for different magnon modes. (b) The conversion efficiency versus  the optomagnonic coupling strength $g_{mb}$ in the Stokes and anti-Stokes processes.  }
	\label{Sfig3}
\end{figure*}

We plot the conversion efficiency, calculated using Eq.~\eqref{conversionefficiency}, versus the external coupling rate $\kappa_a$ in Fig.~\ref{Sfig3}(a) and the optomagnonic coupling strength $g_{mb}$ in Fig.~\ref{Sfig3}(b). It shows that the conversion efficiency increases to a saturation point and then decreases as the external coupling increases. For a relatively weak coupling $g_{mb}$, as in our experiment and others, the conversion efficiencies $\eta_{\rm{as}}$ and $\eta_{s}$ are almost identical, as shown in the inset of Fig.~\ref{Sfig3}(b).  However, the difference between them becomes significant when the optomagnonic coupling $g_{mb}$ increases to the MHz level. This is because the anti-Stokes process absorbs magnons, resulting in the decrease of the magnon excitation number, which in turn restricts the increase of the anti-Stokes scattering probability and the conversion efficiency. In contrast, the Stokes process creates magnons, which can significantly enhance the Stokes scattering probability and thus the conversion efficiency.

\medskip
\textbf{Acknowledgements} \par 
This work is supported by the National Key Research and Development Program of China (No.~2022YFA1405200 and No.~2023YFA1406703), National Natural Science Foundation of China (No.~$92265202$, No.~$11934010$, and No.~$12174329$).

	\medskip
\textbf{Author contributions} \par
Y.P.W. J.L. and J.Q.Y. initiated the research project, W.J.W. designed the experimental system and performed the measurement with input from G.L., Y.P.W. and J.L., W.J.W., Y.P.W., J.L. and J.Q.Y. carried out the data analysis and drafted the manuscript, J.L. and Y.P.W. provided theoretical support, J.Q.Y. supervised the project, and all authors were involved in the discussion of results and the final manuscript editing.

\end{document}